\documentstyle[12pt]{article}
\newcommand{\beq}{\begin{equation}}
\newcommand{\eeq}{\end{equation}}
\newcommand{\beqa}{\begin{eqnarray}}
\newcommand{\eeqa}{\end{eqnarray}}

\begin{document}

\hfill LPT 96-08

\hfill TK 96 09

\hfill nucl-th/9604nnn

\bigskip\bigskip

\begin{center}

{{\large\bf Double neutral pion photoproduction at threshold}}\footnote{Work
supported in part by the Deutsche Forschungsgemeinschaft.}

\end{center}

\vspace{.4in}

\begin{center}
{\large V. Bernard$^a$, N. Kaiser$^b$, Ulf-G. Mei{\ss}ner$^{c \,}$\footnote{
Address after Sept. 1, 1996: FZ J\"ulich, IKP (Theorie), D-52425 J\"ulich,
Germany}}

\bigskip

\bigskip

$^a$Laboratoire de Physique Th\'eorique,
Institut de Physique \\ 3-5 rue de l'Universit\'e, F-67084 Strasbourg
Cedex, France\\and\\
Centre de Recherches Nucl\'eaires, Physique Th\'eorique\\ 
BP 28, F-67037 Strasbourg Cedex 2, France\\ 
{\it email: bernard@crnhp4.in2p3.fr}\\
\vspace{0.3cm}
$^b$Technische Universit\"at M\"unchen, Physik Department T39\\
James-Franck-Stra\ss e, D-85747 Garching, Germany\\ 
{\it email: nkaiser@physik.tu-muenchen.de}\\
\vspace{0.3cm}
$^c$Universit\"at Bonn, Institut f{\"u}r Theoretische Kernphysik\\
Nussallee 14-16, D-53115 Bonn, Germany\\
{\it email: meissner@itkp.uni-bonn.de}\\
\end{center}

\vspace{0.5in}

\thispagestyle{empty}

\begin{abstract}
We consider the chiral expansion of the threshold amplitude for the reaction 
$\gamma p \to \pi^0 \pi^0 p$ to order ${\cal O}(M_\pi^2)$. We substantiate a 
claim that this photoproduction channel is significantly enhanced close to 
threshold due to pion loops. A precise measurement of the corresponding cross 
sections is called for which allows to test chiral perturbation theory.
\end{abstract}

\vfill

\today 

\newpage

\noindent {\bf 1.} Single pion photo-- and electroproduction off
protons have been investigated extensively both from the experimental
and the theoretical sides over the last years. Complementary
information can be gained by the study of double pion production by
real and virtual photons. In Ref.\cite{bkmtwo} we  presented an
exploratory study of double pion photo-- and electroproduction close to
threshold in the framework of heavy baryon chiral perturbation theory. For the
threshold amplitudes, the first two terms in the chiral expansion of
orders one and $M_\pi$, respectively, were given. The most prominent
result of that investigation was the finding that the production cross
section for final states with two neutral pions is considerably enhanced due 
to chiral (pion) loops. In this letter, we wish to sharpen the prediction
for this effect by calculating all next-to-leading order terms of order
$M_\pi^2$ in the threshold amplitude. At this order, one also has to include 
the effects of higher dimension operators. The coefficients of  a subset of 
these have been determined recently from $\pi N$ scattering data in the study 
of the  reaction $\pi N \to \pi \pi N$ \cite{bkmpipin}. Furthermore, 
one has to account for the Roper resonance which plays a very important
role in models that attempt to describe the energy dependence over a 
large range of excitation energies \cite{oset}. Such studies are
complementary to ours which is based on the low energy effective field theory 
of QCD and thus allows to investigate the strictures of the broken chiral 
symmetry for small external momenta, i.e. it is valid exclusively in the
threshold region. The existing data of the two-pion photoproduction reactions
from DAPHNE \cite{daphne} and TAPS \cite{taps} are not yet conclusive for the
threshold region. However, recently new precision data have been taken
with TAPS \cite{stroh} for the reaction $\gamma p \to \pi^0 \pi^0 p$ also close
to threshold which will be analysed in the near future. An improved calculation
is therefore called for  to allow new tests of chiral perturbation theory in
this particular  low energy reaction.   

\medskip

\noindent {\bf 2.} First, we have to give some necessary definitions. The 
threshold matrix element for double neutral pion photoproduction off protons 
in the center of mass frame  takes the form
\begin{equation} 
T(\gamma p \to \pi^0\pi^0 p)^{\rm thr}_{\rm cm} = i \vec \sigma
\cdot ( \vec \epsilon \times \vec k) \, {\cal M}_{00} \, \, , 
\label{me}
\end{equation}
where $\vec \sigma$, $\vec \epsilon$ and $\vec k$ denote the spin of
the nucleon, the photon polarization vector and the photon three--momentum (in
 the cm frame), in order, subject to the transversality constraint $\vec 
\epsilon \, \cdot \vec k \, = 0$. The (complex) threshold amplitude ${\cal
M}_{00}$ encodes the physics. Assuming the photoproduction amplitude to be
constant in  the threshold region and approximating the three--body phase space
and photon flux by a simple analytic formula\footnote{This approximation has
been shown to  work excellently in \cite{bkmtwo}.},
the  unpolarized total cross section near threshold is
\begin{equation} 
\sigma_{\rm tot}(E_\gamma) = {M_\pi^2 (1+\mu)\over 64 \pi^2
(1+2\mu)^{11/2}} \,|{\cal M}_{00}|^2\,\cdot 
(E_\gamma - E_\gamma^{\rm thr})^2 \, \, \, , 
\label{xsec}
\end{equation}
with $E_\gamma$ the photon lab energy and $\mu = M_\pi /m =0.144$
the ratio of the (neutral) pion to the proton mass. The value of the
threshold photon energy in the lab frame is $E_\gamma^{\rm thr} = 2
M_{\pi^0}(1+\mu) = 308.8\,$MeV. 

\medskip

\noindent {\bf 3.} To calculate the chiral expansion of the threshold 
amplitude ${\cal M}_{00}$, we make use of heavy baryon chiral
perturbation theory \cite{jm,bkkm}. Our notation is identical to the
one used in Ref.\cite{bkmtwo} and we refer to that reference for further
details. We will extend the calculation of that paper by working out
all terms of order $q^4$ in the Lagrangian, which amounts to tree
graphs with insertion from the dimension one, two, three and four effective
pion--nucleon Lagrangian as well as pion loop graphs with at most one
dimension two insertion. Since the quantities $\vec \epsilon$ and 
$\vec k$ count as small momenta, we therefore get all terms 
up--to--and--including ${\cal O}(M_\pi^2)$ in the chiral expansion 
of ${\cal M}_{00}$.

Let us now consider the various contributions. The tree graphs (sequential
photon absorption and pion emissions on the proton line) with the lowest
order couplings together with the higher order magnetic moment terms
are conventionally called the Born terms. The Born contribution to the 
threshold amplitude is easily evaluated (notice that for double neutral pion
production the first non--vanishing Born term starts at order $M_\pi$), 
\begin{equation} 
{\cal M}_{00}^{\rm Born} = {eg_A^2 M_\pi \over 2 m^2 F_\pi^2}
\biggl[ 1 - {M_\pi \over 4m } (1+3\kappa_p)\biggr] = (4.9 -1.2 )\, {\rm
GeV}^{-3} = 3.7 \, {\rm GeV}^{-3} \, \,  \label{born} \end{equation}
using $ e^2/ 4\pi = 1/137.036$, 
$F_\pi = 92.4\,$MeV, $M_\pi =139.57\,$MeV,
$m=938.27\,$MeV, $\kappa_p = 1.793$ and $g_A = 1.32$ as determined from the
Goldberger--Treiman relation (with $g_{\pi N} = 13.3$). Isospin breaking
via the difference of the neutral and charged pion masses is negligible here
($3\%$-effect). The one-pion loop diagrams contributing  at order $q^3$ are 
taken from \cite{bkmtwo},
\begin{equation} 
{\cal M}_{00}^{\rm q^3-loop} = {3eg_A^2 M_\pi \over 128 \pi
F_\pi^4} \biggl\{2+{\pi\over2} +i\,\bigl[2\sqrt{3} - \ln(2+\sqrt{3})\bigr]
\biggr\}= (26.9+16.2\, i )\, {\rm GeV}^{-3} \, \, ,
\label{loop3}
\end{equation}
where the imaginary part is related to the rescattering type of
diagrams which have a right--hand cut starting at the single pion production
threshold, $s=(M_\pi+m)^2$, which is well below the two pion production 
threshold. Notice that the ${\cal O}(q^3)$ loop contribution Eq.(\ref{loop3})
is much larger than the one from the Born terms, Eq.(\ref{born}). This is,
indeed, the large chiral loop effect which enhances the cross section
for the photoproduction of two neutral pions off protons considerably.
At this order, one could also have $1/m$ suppressed tree graph contributions
proportional to the low--energy constants $c_i$. As shown in
\cite{bkmtwo}, these vanish due to the threshold selection rules or
cancel pairwise. However, at one order higher, one has such
contributions. In the notation of Ref.\cite{bkmpipin}, the terms
proportional to the low energy constants $c_i$ read
\begin{equation} {\cal M}_{00}^{c_i} = {e M_\pi^2 \over m^2 F_\pi^2}
\biggl[-2c_1 +(1+\kappa_p)\,c_2'+(1+2\kappa_p)\,c_2''-c_3\biggr] =
18.4 \, {\rm GeV}^{-3} \, \, ,
\label{ci}
\end{equation}
using the central values of the $c_i$ as determined in \cite{bkmpipin} from a 
fit to the subthreshold pion--nucleon scattering amplitudes, the $\pi N$
$\sigma$-term and some $\pi N$ P-wave scattering volumes ($c_1 = -0.64$,
$c_2'=-5.63$, $c_2''=7.41$, $c_3=-3.90$, all in GeV$^{-1}$). In a microscopic 
picture, these tree terms $\sim c_i$ subsume all s--, t-- and
u--channel resonance excitations in the $\pi^0 p$ scattering amplitude.
Combining these with photon absorption on the proton line, one is
lead to a contribution to $\gamma p \to \pi^0 \pi^0p$, i.e. the 
expression Eq.(\ref{ci}) for double neutral pion production off
protons. The anatomy of such a chiral coefficient is discussed in 
\cite{ulfmit}, see also \cite{bkmr}. We note that contributions from single
$\pi^0$--photoproduction via resonance exchanges (see Ref.\cite{bkmpi0}) 
followed by $\pi^0$ emission off the proton line vanish to the order we are
working here. 

Next, we consider the pion loops at order $q^4$. Many of these are similar
in structure to the ones evaluated in \cite{bkmpipin} for the reaction
$\pi N \to \pi \pi N$. Furthermore, one can show that the sum of all one-loop
diagrams with one insertion proprotional to $c_i$ $(i=1,2,3)$ vanishes. 
Consequently, only the dimension two operators with the LEC $c_4$ (central
value $c_4= 2.25$ GeV$^{-1}$ \cite{bkmpipin}) together with the anomalous 
magnetic moments of the proton ($\kappa_p$) and the neutron ($\kappa_n= 
-1.913$) and $1/m$ suppressed vertices give a contribution to the $q^4$ one
loop result  
\begin{eqnarray} 
{\cal M}_{00}^{\rm q^4-loop} = && {eM_\pi^2 \over 64\pi^2 m
F_\pi^4} \biggl\{2\bigl[\kappa_p(11g_A^2-1)+\kappa_n(3g_A^2+1)+20g_A^2-12mc_4-
4\bigr] \ln{M_\pi \over \lambda} \nonumber \\ && - (29g_A^2+24mc_4 +6){\pi^2
\over 16} +g_A^2 \pi +3(4g_A^2-4m c_4-1) \sqrt{2}\ln(1+\sqrt{2}) \nonumber \\ 
&& - 3(g_A^2+2m c_4+\frac{1}{2} )\ln^2(1+\sqrt{2})+g_A^2 (13+11\kappa_p+3
\kappa_n) \sqrt{3}\ln(2+\sqrt{3}) \nonumber \\ && -\frac{3}{4}g_A^2
\ln^2(2+\sqrt{3}) +7+24m c_4+(g_A^2-1)\kappa_n-\frac{g_A^2}{3}
(37+31 \kappa_p) + \kappa_p \nonumber \\ & & - i \, \frac{\pi}{2}g_A^2
\biggl[(9+11 \kappa_p + 3 \kappa_n) \sqrt{3} + \frac{11}{2} \ln(2+\sqrt{3})
\biggr] \biggr\} \nonumber \\ & &= (-10.3 -17.6\,i)\, {\rm GeV}^{-3}
\label{loop4}
\end{eqnarray}
Here, $\lambda$ is the scale of dimensional regularization. In what
follows, we set $\lambda = m$. The (spurious) scale dependence is rather mild,
for $\lambda = m^* = 1.44$ GeV, the real part in Eq.(\ref{loop4}) changes to 
$-13.3$ GeV$^{-3}$. We note that in contrast to the $q^3$ calculation, the
loops are  no longer finite. Furthermore, the imaginary part at orders $q^3$
and $q^4$ are of the same size but of opposite sign. These first two terms in
the chiral expansion of Im~${\cal M}_{00}$ therefore do not give a reliable
estimate of the imaginary part. To get a better handle on it, we make use of
the Fermi--Watson final--state theorem. Unitarity relates the imaginary part
Im~${\cal M}_{00}$ to the rescattering process $\gamma p \to \pi N \to \pi^0
\pi^0 p$ with intermediate $\pi N$ states on mass-shell. The resulting
exact expression is: 
\begin{eqnarray} 
{\rm Im}~{\cal M}_{00} &=& - {3 M_\pi (2+3\mu)(2+\mu)\over 8
(1+2\mu )( 1 + \mu) } \biggl[ \sqrt{2}\, D_2^*\,  M_{1-}^{\pi^+ n}
+(2D^*_1+D^*_2) \, M_{1-}^{\pi^0 p} \biggr] \nonumber \\ &=& {3 M_\pi
(2+3\mu)(2+\mu)\over 8 (1+2\mu )( 1 + \mu) } \biggl[ {\cal A}^*_{10}
\,\,{_pM}_{1-}^{(1/2)} -{2\over 15} \sqrt{10}\, {\cal A}_{32}^*\,\,
M_{1-}^{(3/2)} \biggr]\simeq 7 \, {\rm GeV}^{-3}  \label{imag} \end{eqnarray}
where the factor in front of the brackets is 
the ratio of the intermediate pion momentum squared to the photon momentum,
$|\vec q_\pi|^2/|\vec k|$ (in the center-of-mass system). $D_{1,2}$ are the
threshold amplitudes for the reaction $\pi N \to \pi \pi N$ \cite{bkmpipin} 
and the $M_{1-}$ are the pertinent single pion photoproduction P--wave magnetic
dipole amplitudes. In the second line of Eq.(\ref{imag}), the decomposition
into channels with definite total isospin 1/2 and 3/2, respectively, of the
intermediate $\pi N$ system is given. The threshold $\pi\pi N$
amplitudes have the (central) values \cite{bkmpipin}  \begin{equation} 
D_1 = 2.26\,{\rm fm}^3, \quad D_2 = -9.05 \,{\rm fm}^3, \quad
{\cal A}_{10} = 8.01\,M_\pi^{-3}, \quad {\cal A}_{32}= 2.53\,M_\pi^{-3}
\, \, , \label{valpipin} \end{equation} 
and for the photoproduction magnetic multipoles we take the values from the
analysis of  Berends and Donnachie \cite{berends} at $E_\gamma = 310...320$
MeV (see also Ref.\cite{pfeil}),  
\begin{equation} 
_pM_{1-}^{(1/2)} \simeq 1\cdot 10^{-3}\,M_\pi^{-1} , \quad
M_{1-}^{(3/2)} \simeq -6\cdot 10^{-3}\,M_\pi^{-1} \, \, , 
\label{valpwave}
\end{equation}
which leads to the value of Im~${\cal M}_{00}$ given in Eq.(\ref{imag}). The
imaginary parts of the (actually complex) $\pi\pi N$ threshold and pion
photoproduction magnetic dipole amplitudes are negligible for our estimate in 
Eq.(7), since the respective $\pi N$-phases $(\delta_{11} \simeq 2^\circ, \, 
\delta_{31} \simeq -4^\circ$) are tiny. Even though the imaginary part
Im~${\cal M}_{00}$ calculated to order $q^4$ is numerically not reliable, it
serves as an important check on the evaluation of the loop integrals.    

Finally, we have to account for tree level insertions from the
dimension three and four local contact terms. As detailed in
Ref.\cite{bkmtwo}, we make use of the resonance saturation hypothesis
to estimate these. As argued before, single $\Delta$ excitation (see
Fig. 5a--h in \cite{bkmtwo}) is to order $q^4$ subsumed in the tree terms
$\sim  c_i$,  Eq.(\ref{ci}), so we are left with double $\Delta$ graphs like in
Fig. 5i,j,k in \cite{bkmtwo}.
To proceed, we generate the $\Delta \Delta \gamma$ coupling by minimal 
substitution in the free $\Delta$ Lagrangian and the unknown $\Delta
\Delta \pi$ coupling with help of the spin--3/2 axial current.
Using the coupling constant relation  $g_{\pi \Delta\Delta} = 
4 \, g_{\pi N} \, / 5 $ from the SU(4) quark model, we get
\begin{eqnarray} 
{\cal M}_{00}^{\Delta\Delta} &=& {e g_A^2 M_\pi^2\over 2 m
m_\Delta^2 F_\pi^2 }\biggl\{{4 \over 15}\sqrt{2}g_1(1+2Y)(1-2Z) +{m_\Delta
\over m-m_\Delta} +{m^2 \over 3 m_\Delta^2}(1-2Z)^2 \nonumber \\ && +{m\over 3
m_\Delta}(1-2Z)(5+2Z) +{7\over 4} - Z - Z^2\biggr\} = 1.2 \,{\rm GeV}^{-3}
\label{DD}
\end{eqnarray} 
where the first term related to the graphs Fig. 5i,k in \cite{bkmtwo} (the
photon does not couple to two $\Delta^+$) dominates. The off--shell parameters
$Y$ and $Z$ have been determined in \cite{bkmpi0} ($g_1=5,\,Y=0.1,\,Z=-0.2$). 
Inserting these, we find that the double $\Delta$ contribution is negligible. 
This agrees with the findings of Bolokhov et al. \cite{bolo} for the reaction 
$\pi N \to \pi \pi N$. We remark that such terms are exactly zero in the 
non--relativistic limit (static isobar model) \cite{oset}. Another possible
contribution comes from loop diagrams where the photon couples to a virtual
$K^+$ meson and $\Sigma^0$ and $\Lambda$ hyperons appear in the intermediate
state. The respective calculation in chiral $SU(3)$ gives after expanding in
inverse powers of the kaon mass
\begin{equation} {\cal M}_{00}^{\rm K^+-loop} = { e (D^2 +3F^2) M_\pi^2 \over
48 \pi F_\pi^4 M_K} = 1.4\, {\rm GeV}^{-3} \,\, \label{kal} \end{equation}
using $M_K = 493.6$ MeV and $D=0.76$, $F=0.50$ for the axial vector coupling
constants. As in single pion photoproduction \cite{sven} \cite{bkmpi0} 
the (frozen) $K^+$--loop effects are negligibly small.

The largest
resonance contribution at order $M_\pi^2$ comes indeed from the Roper 
resonance $N^*(1440)$. To be specific, the pertinent Feynman diagrams involve 
the $N^* N \pi \pi$ S-wave vertex which is discussed in detail in
Ref.\cite{bkmpipin}. To leading order, it is parametrized in terms of two 
coupling constants, denoted $c_1^*$ and $c_2^*$, respectively. We also need the
$N^{*+} p \gamma$ coupling, $\kappa^*$, defined via the relativistic vertex
$(e \kappa^*/ 4m) \bar \Psi \sigma_{\mu\nu} F^{\mu\nu} \Psi^*$. The 
transition magnetic moment $\kappa^*$ can be determined from the Roper 
radiative  decay width,
\begin{equation} 
\Gamma(N^{*+}\to p \gamma) = {e^2 \kappa^{*2} k_\gamma^3
\over 4 \pi m^2} = {k_\gamma^2 m \over \pi m^*} |A_{1/2}|^2 = 0.18 \,
{\rm MeV}, \quad \kappa^* = -0.56 \, \, \label{nstar}
\end{equation}
using the value $A_{1/2} = - 0.072$ GeV$^{-1/2}$ from the latest PDG tables and
$k_\gamma = (m^{*2} - m^2)/2m^* = 414$ MeV. The strong couplings
constants $c_{1,2}^*$ are only known within some broad ranges. We therefore
give an upper limit for the Roper contribution using the maximal value 
of $c_1^*+c_2^*= -4.9$ GeV$^{-1}$ \cite{bkmpipin} 
\begin{equation} 
{\cal M}_{00}^{N^*} = {2eM_\pi^2 \,\kappa^*(c_1^*+c_2^*)
\over m(m^*-m) F_\pi^2} < 11.7\, {\rm GeV}^{-3} \, \,.\label{rop}\end{equation}
The value ${\cal M}_{00}^{N^*}= 11.7$ GeV$^{-3}$ in Eq.(\ref{rop}) also 
includes the possible enhancement factor $[1-4M_\pi^2/(m^*-m)^2]^{-1}= 1.45$
which occurs if the energy denominators are not chirally expanded.\footnote{We
are grateful to Eulogio Oset for pointing this out.}  If one
chooses e.g. the central values of the Roper couplings $c_1^*+c_2^* = -1.6$
GeV$^{-1}$ as in \cite{bkmpipin} this contribution is strongly reduced to
${\cal M}_{00}^{N^*} = 2.6$ GeV$^{-3}$. Obviously, the Roper contribution has
presently the largest theoretical uncertainty. Higher resonances play therefore
no role for the threshold amplitude ${\cal M}_{00}$, i.e. their contribution is
well within the theoretical uncertainty.

\medskip

\noindent {\bf 4.} We are now in the position to calculate the total
cross section for $\gamma p \to \pi^0 \pi^0 p$ in the threshold
region. Consider first the real part of ${\cal M}_{00}$. We find that
the leading contribution of order $M_\pi$ is at least a factor 1.5 larger than
the first correction at order $M_\pi^2$, Re~${\cal M}_{00}^{q^3} = 31.8
\,$GeV$^{-3}$ and for the maximal value of the Roper contribution, 
Re~${\cal M}_{00}^{q^4} = 21.2 \,$GeV$^{-3}$, in order. For the central Roper
couplings of \cite{bkmpipin} the order $M_\pi^2$ correction is even further 
reduced to Re~${\cal M}_{00}^{q^4} = 12.1 \,$GeV$^{-3}$, which shows that the
chiral expansion is well--behaved, i.e. the next--to--leading order correction
is smaller than the leading term. For  the maximal $N^*(1440)$ contribution 
together with the empirical  imaginary part, we have
\begin{equation}
{\cal M}_{00} \simeq (53 + 7\,i)\,{\rm GeV}^{-3} \, \, \, ,
\end{equation}
which shows that the imaginary part is not important for the total cross 
section, i.e. it amounts to a $2\%$--effect. Altogether, the predicted near 
threshold cross section is then\footnote{We use the neutral pion mass
in the phase space factor as discussed in \cite{bkmtwo}.}
\begin{equation}
 \sigma_{\rm tot}(E_\gamma) \le 0.91 \, {\rm nb} \, \biggl(
{E_\gamma - E_\gamma^{\rm thr} \over 10\, {\rm MeV}}
 \biggr)^2 \,, \qquad E_\gamma^{\rm thr} = 308.8\, {\rm MeV} \, \, . 
\label{cross}
\end{equation}
For comparison,  at order $q^3$, the prefactor in Eq.(\ref{cross}) is
0.41~nb. We stress that the 0.91~nb given in Eq.(\ref{cross}) are the
upper limit. For the central Roper couplings in \cite{bkmpipin} which lead to a
very good description of the $\pi\pi N$ threshold amplitude $D_2$, this number
is reduced to 0.63~nb.

In summary, we have evaluated all next--to--leading order corrections to the
threshold amplitude for the reaction $\gamma p \to \pi^0 \pi^0 p $. The
dominant contribution still comes from the chiral loops at leading
(non-vanishing) order, $q^3$. The higher order $q^4$ corrections are in total 
not very large and increase the cross section by about a factor 1.6 to 2.2 with
the uncertainty mainly related to the size of Roper resonance contribution. In
order to test the prediction Eq.(\ref{cross}) a precise measurement of these
near threshold cross sections is now called for.


\section*{Acknowledgements}

One of us (UGM) thanks the Institute for Nuclear Theory at the University of
Washington for its hospitality and the Department of Energy for partial support
during the completion of this work.  


\end{document}